# Diffusion stabilizes cavity solitons in bidirectional lasers

**Isabel Pérez-Arjona[(1)], Víctor Sánchez-Morcillo[(1)], Javier Redondo[(1)], Kestutis Staliunas[(2)], and Eugenio Roldán[(3)]**

[(1)]*Dept. de Física Aplicada, Escola Politècnica Superior de Gandia, Universitat Politècnica de València, Ctra. Nazaret-Oliva S/N, 46730-Grau de Gandia, Spain*

[(2)]*Dept. de Física i Enginyeria Nuclear, Universitat Politècnica de Catalunya, Colom 11, 08080-Terrassa, Spain*

[(3)]*Dept. d'Òptica, Universitat de València, Dr. Moliner 50, 46100-Burjassot, Spain*

**Abstract:** We study the influence of field diffusion on the spatial localized structures (cavity solitons) recently predicted in bidirectional lasers. We find twofold positive role of the diffusion: 1) it increases the stability range of the individual (isolated) solitons; 2) it reduces the long-range interaction between the cavity solitons. Latter allows the independent manipulation (writing and erasing) of individual cavity solitons.

## References and links

## 1. Introduction

Spatially extended nonlinear systems, when brought far enough from the thermodynamic equilibrium, can display a rich variety of dissipative structures or patterns [1]. Nonlinear optical cavities, such as lasers or photorefractive oscillators, belong to this class of pattern forming systems [2,3] when the cavity Fresnel number is large enough (wide aperture cavity) for sustaining a large number of transverse cavity modes. In optical systems, dissipative structures appear in the plane orthogonal to the cavity axis, i.e., they are two dimensional structures, hence the name transverse patterns, although three-dimensional patterns have also been predicted [2,3]. Among the variety of possible transverse patterns, localized structures (LSs) are of particular interest because of their potential for information processing purposes [4].

Recently, LSs in bidirectional lasers have been predicted [5] and attracted interest. A bidirectional laser is a ring cavity laser in which emission is allowed to occur in two possible directions, say clockwise and counterclockwise. The LSs in this laser type were shown in the simplest model of wide aperture bidirectional lasers, consisting of two coupled Ginzburg-Landau equations [5], see Eqs. (1) below, for slightly anisotropic laser cavities (anisotropic in the sense that the cavity losses for the two counter propagating fields were different in a few per cent). This prediction is relevant because "pure" lasers do not exhibit LSs (except optical vortices) unless some modifications of nonlinearities, and/or unless the external actions. Such modified systems are lasers with intracavity saturable absorbers [6] or intracavity (Kerr) nonlinearity [7], lasers with a cw [8] or a periodic injection [9,10] ("rocked" lasers), or lasers with frequency-selective optical feedback [11].

The existence of the LSs reported in [5] is based on the fact that a plane-wave laser cannot sustain stable bidirectional cw emission, but rather unidirectional cw emission in either of two possible emission directions, as both field components compete for a common population inversion [12]. Bidirectional emission is then possible only in a pulsed manner, i.e. in the form of periodic or chaotic alternation of unidirectional emission [12]. However, when a plane-mirrors large aspect ratio cavity is considered (i.e., a cavity that allows the oscillation of, in principle, any transverse mode), the above mentioned restriction on bidirectionality holds only locally. This means that the emission in the clockwise direction occurs in a given region of the plane transverse to the cavity axis, which we take to be the **z**-axis, whereas the emission in the counterclockwise direction takes part in an adjacent transverse region. The continuity of the solutions between these two regions (domains) in the transverse plane implies the existence of a front connecting the two solutions. Fronts are highly unstable in general as they tend to move, which leads to eventual unidirectional motion in the entire transverse plane. However, as shown in [5], if the two intracavity fields have slightly different losses, two of the fronts approach each other and can lock forming stable LSs. These LSs look like bright LSs in the weak field (the field with more loses) while in the strong field they resemble dark LSs, as a dip appears in the otherwise homogeneous (bright) solution. This nomenclature should not hid the fact that these LSs do not arise as the result of bistability between a pattern and the homogeneous null solution (which is the case of usual bright LSs) or to phase bistability (which is the case of usual dark LSs). The origin of the bidirectional laser LSs lies on the bistability between the two propagation directions and is the result of a fine balance between the tendencies of fronts to move in opposite directions. The balance is due to the presence of sinks and sources of traveling waves in between the two fronts.

In [5] these LSs were considered to be cavity solitons (CSs). But as remarked in [13] the LSs must verify a number of requirements in order to be considered as CSs, and not all of them were verified in the bidirectional laser model of [5]. More specifically, the above described LSs [5] cannot be written/erased without disturbing the other neighboring LSs, which violates the basic requirement for the information storage and manipulation. In [13] a solution for this drawback was proposed using an injection of a weak signal in one or the two counter-propagating fields. The weak injection fixes the intracavity fields phases what prevents the appearance of traveling waves in the transverse space (equivalently tilted waves in a three dimensional representation). This eventually suppresses the long-scale interaction between LSs. In this way the LSs are no more affected when neighboring LSs are written or erased, and then they belong to the narrower class of CSs [13]. In the present letter we argue that transverse diffusion can also stabilize the LSs of the bidirectional laser in such a way that they behave as true CSs. In this way, we show that the bidirectional ring laser continues being a rare laser that supports CSs without the need of any additional external action. Moreover, we show that diffusion stabilizes individual LSs in the sense that their stability area in the parameter space becomes larger in the presence of diffusion. We remark that with diffusion the LSs also become stable for the isotropic case (when both unidirectional waves have the same losses), which is also a serious advantage, from the practical point of view.

## 2. Bidirectional ring laser model.

In [5] it was shown that a wide-aperture class-A bidirectional laser (for which all material variables relax much faster than the intracavity field), working close to the emission threshold, can be modeled by the following coupled Ginzburg-Landau equations:

$$\partial_\tau F_1 = (A - 1 - |F_1|^2 - 2|F_2|^2)F_1 + (d + i)\partial_\xi^{\,2} F_1, \tag{1a}$$

$$\partial_\tau F_2 = (A - \sigma - |F_2|^2 - 2|F_1|^2)F_2 + (d + i)\partial_\xi^{\,2} F_2, \tag{1b}$$

where $F_i(\xi,\tau)$ are proportional to the intracavity field amplitudes of the two counter-propagating fields, $A$ accounts for the pump strength, $\sigma = \kappa_2/\kappa_1$ with $\kappa_i$ the decay rate of field $F_i$ (we take $\sigma \geq 1$ without loss of generality, hence $F_1(F_2)$ is the strong (weak) field), $\tau = \kappa_1 t$, and $\xi$ is the transverse coordinate normalized to the square of the diffraction coefficient.

In [5] only the case $d = 0$ was considered. Here we also account for the diffusion term $d\partial_\xi^{\,2} F_2$ introduced phenomenologically ($d$ is the diffusion coefficient normalized to diffraction) and study its influence on the stability and dynamics of LSs. There are several origins of the transverse diffusion: on one hand, the transverse diffusion is always present on any nonlinear optical cavity because of diffraction losses in finite aperture cavity, and these are properly modeled with the diffusion term [14]. On the other hand, it can be rigorously shown, that for large aperture class-C lasers the appropriate model includes a Swift-Hohenberg term of the form $(\delta + \partial_\xi^{\,2})^2 F_i$ ($\delta$ is a detuning) that works as a "superdiffusion" [2,3]. As we keep our study as general as possible, we study the simplest Eqs. (1), but we have in mind that from a qualitative point of view we are also covering, in a sense, the case of class-C lasers. Next we pass to describe our numerical findings.

## 3. Numerical results.

We have numerically integrated Eqs. (1) by means of a split-step method on a grid of 512 spatial points and with the total integration length L=316. We have considered a flat-top profile for the pumping $A$ (modeled with a super-Gaussian function) in order to take into account the finite transverse extension of the laser excitation, similarly to [5].

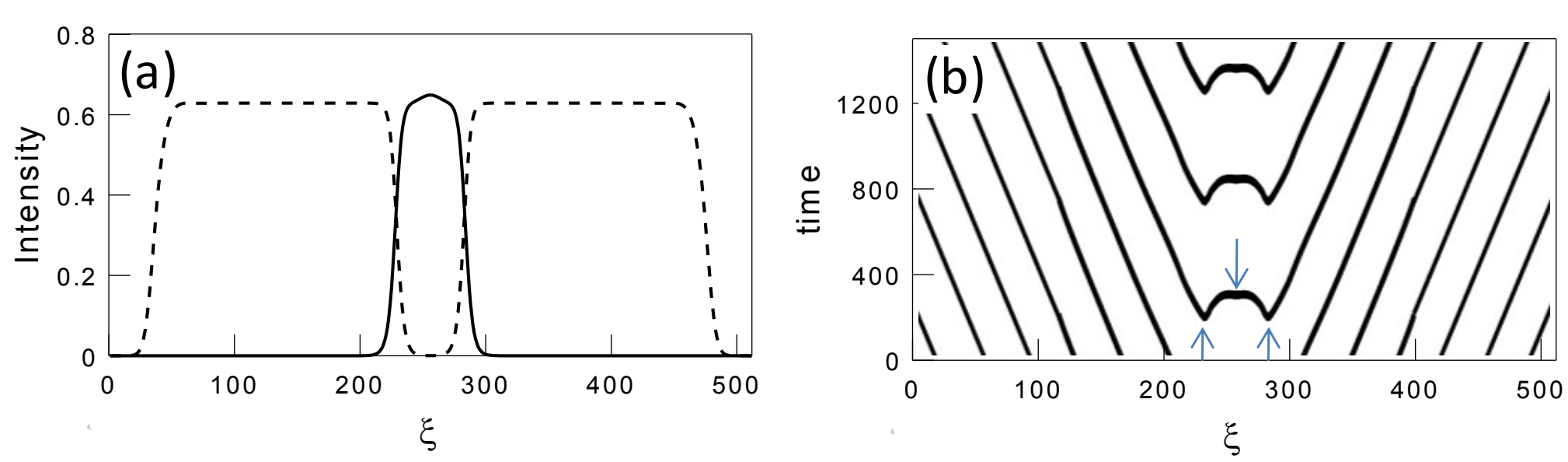

**Fig.1**. Bidirectional laser localized structure as obtained from Eqs. (1) for $\sigma = 1$ (isotropic case), $d = 0.3$ and $A = 1.4$. In (a) the dashed (full) lines correspond to the transverse intensity distribution for the strong (weak) field. In (b) the phase is shown as a function of time. Notice the presence of sources and sinks of traveling waves in the transverse dimension, indicated by arrows.

Fig. 1(a) shows the intensity profile of a stationary LS obtained for an isotropic cavity ($\sigma = 1$), for $A = 1.4$ and $d = 0.3$. The two counter-propagating fields are represented by continuous and dashed lines respectively. In Fig. 1(b) the time evolution of the corresponding phase profile is shown. The curves represent equiphase contours, where the tilted, straight lines denote the emission of traveling waves from the LSs, whose relevance will be considered below. We note that the existence of the stationary LSs for $\sigma = 1$ is a new result made possible by the introduction of diffusion.

In Fig. 2 we represent the domain of existence, in the plane $<\sigma,d>$, of the LSs for a fixed value of the incoherent pump $A = 1.4$. Notice that the case $d = 0$ was already treated in [5] where for $\sigma \to 1$ a Hopf bifurcation has been identified which makes the LSs height and width oscillate in time. Moreover, for $\sigma = 1$ and $d = 0$ no proper LSs exist as their width oscillates between a minimum and a maximum that extends over the whole spatial extension [5]. Fig. 2 shows that the influence of diffusion is the following. On one hand, it shrinks the domain of the self-pulsing LSs till it disappears for $d \approx 0.1$, allowing for the existence of steady LSs even in the isotropic case $\sigma = 1$. On the other hand, the diffusion first enhances the domain of existence of LSs, and then shrinks it, but quite weakly. In conclusion, from the view point of the domain of existence of LSs, small diffusion (say $0.1 < d < 0.5$) plays a positive role for the case shown in Fig. 2. Thus diffusion is advantageous for the individual LSs as they become more stable (the pulsing domain shrinks and then disappears) and can exist for a perfectly isotropic cavity.

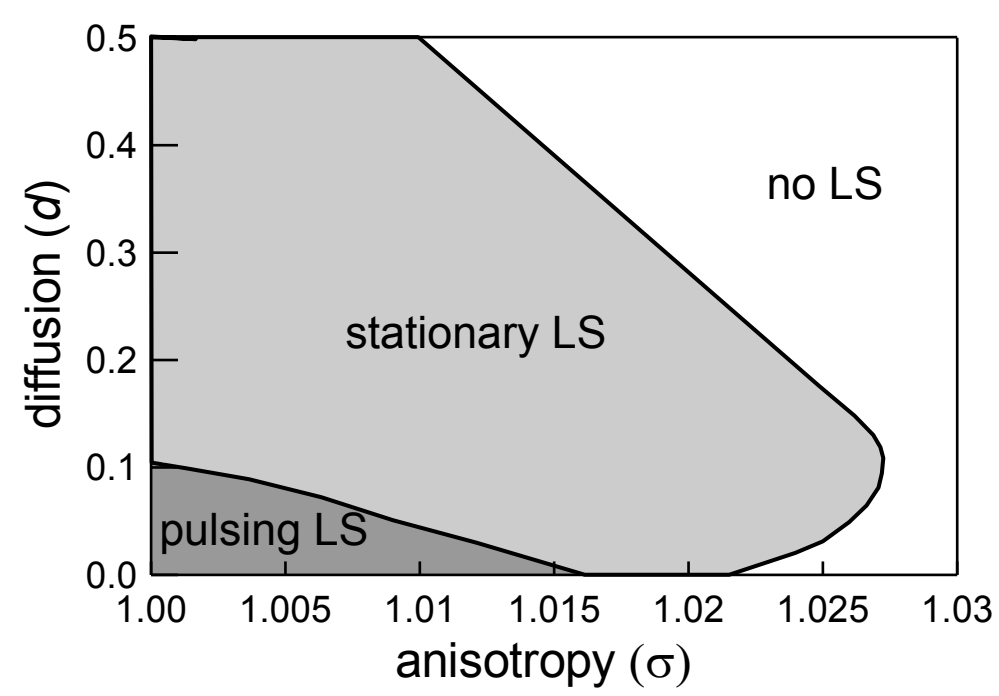

**Fig. 2**. Domain of existence of the LSs in the plane $<\sigma,d>$ for a pump value $A = 1.4$. Shaded areas correspond to the regions the stable LSs are found.

Next we show the main result of the present letter namely that the LSs become true CSs for appropriate diffusion values. With that aim, we investigate the dynamics of a cluster of LSs in the presence of diffusion. Fig. 3 shows the evolution of a cluster of five LSs which are injected at different times: the central LS is injected at $\tau = 0$, the rest are injected later. The panels of the upper row correspond to five different (increasing) values of $d$ for $\sigma = 1.01$, while the five panels of the lower row correspond again to five different (increasing) values of $d$ for $\sigma = 1.005$, see figure caption. The bright LS corresponding to the weak field is depicted in the figure.

It is evident that for the lower values of $d$ the injected LSs do not survive [cases (a)-(c) in Fig.(3)]. This is because they are advected by the traveling waves emitted by the sources located at the central LS (see Fig. 1(b)). As the diffusion coefficient is increased, the injected LSs survive longer, and for large enough $d$ [cases (d) and (e) in Fig.(3)] they remain. Note also that, in this parameter region, some of the LSs can be erased, without affecting significantly the other ones in the array. In this sense, as stated in the introduction, they can be considered as true CSs. However, when $d$ is too large, a strong interaction between the different LSs is again apparent. We can conclude that the optimum values of $d$ correspond to those in the central region of the stability domain of LSs, see Fig. 2.

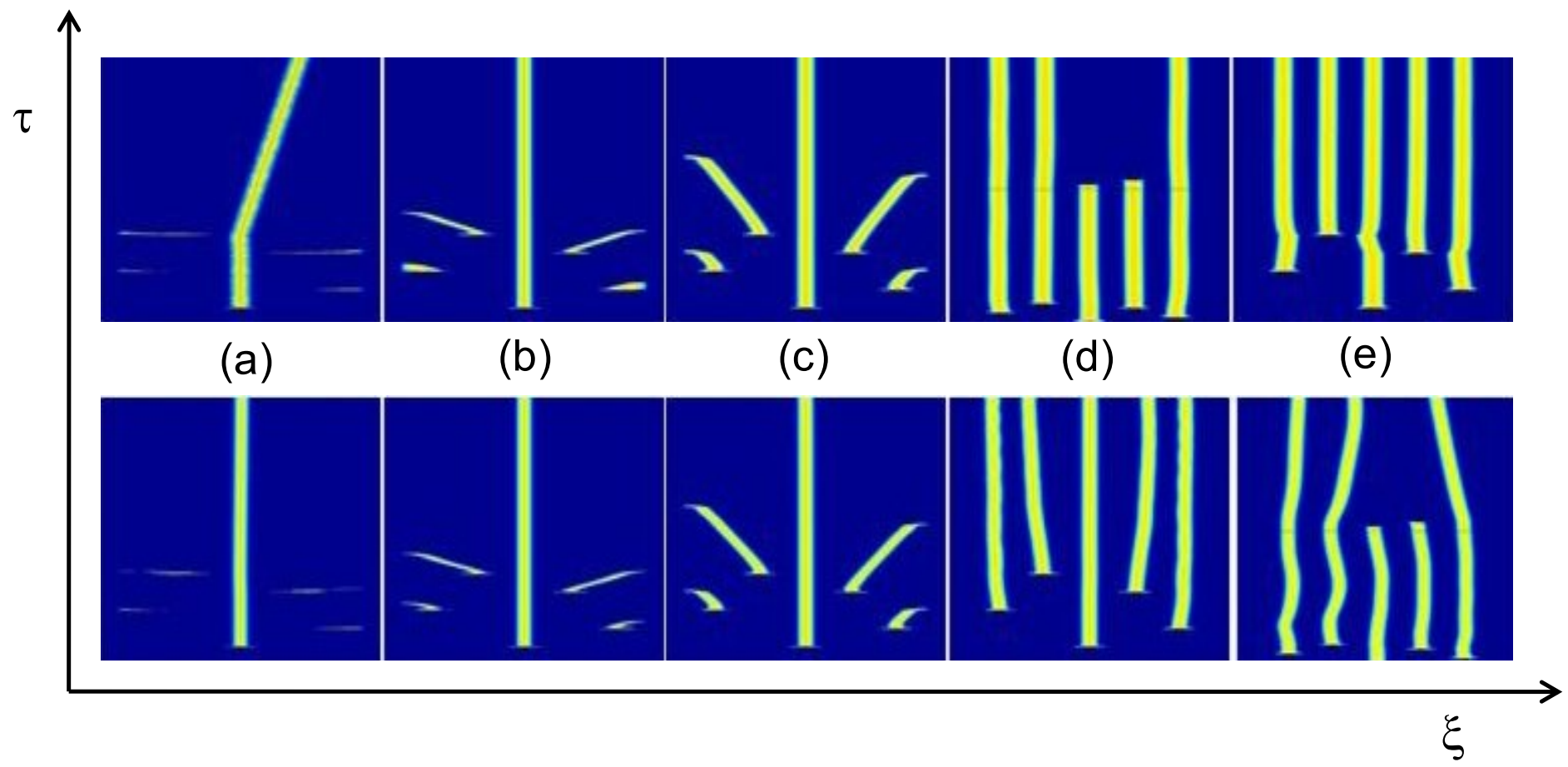

**Fig.3.** Evolution of a cluster of five LSs for σ = 1.01 (Upper row) and σ = 1.005 (Lower row), for different values of the diffusion coefficient. From left to right: $d = 0$ (a), $d = 0.1$ (b), $d = 0.2$ (c), $d = 0.3$ (d), and $d = 0.4$ (e). Time span is 2500 in all Figures except in those showing the erasing of LSs, where a time span of 5000 was chosen for clarity.

The origin of the stabilization introduced by diffusion can be traced back to the dynamics of the traveling waves emitted by the sources. The CS is a source of two traveling waves propagating outwards from the soliton. (More precisely the CS consists of two sources and one sink, see Fig. 1(b), but in overall it behaves as a source from a long distance perspective.) In the absence of diffusion, the traveling waves emitted by the CSs interact with the neighboring ones, pushing them apart, and eventually destroying them [see, e.g., Fig. 3(a)-(c)]. The increasing diffusion weakens the effect of these traveling waves as their velocity (equivalently the tilt angle of the tilted wave) reduces with increasing $d$. The effect resembles the dependence of the tilted waves emitted by sources in the usual complex Ginzburg-Landau Equation [15] where, as shown analytically, the velocity of the traveling waves decreases with increasing diffusion. The situation here is more involved, and the analytical evaluation is more complicated, therefore we calculated numerically the phase velocity of the traveling wave as emitted by isolated LS. Fig.4 shows the quick decrease of the phase velocity of the tilted wave with increasing diffusion, for different values of the anisotropy σ, and thus qualitatively explains the numerically observed effect.

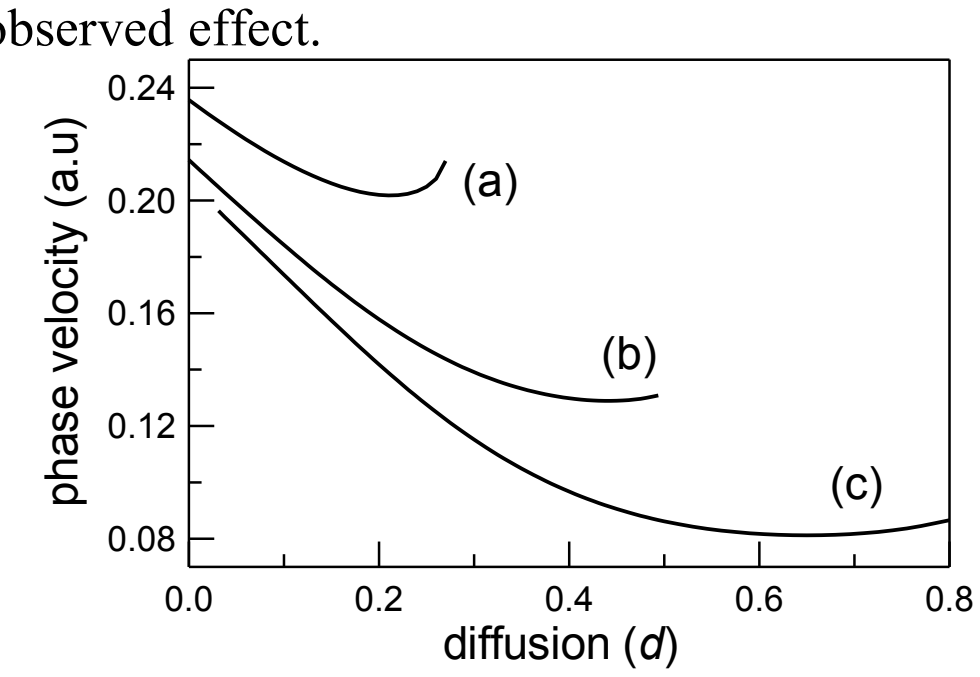

**Fig. 4.** Phase velocity as a function of diffusion $d$ for σ = 1.020 (a), 1.010 (b), 1.005 (c). The velocity is calculated by evaluating the derivative of the phase with respect the time, at a point sufficiently far from the center of the LS.

## 4. Conclusions

We have numerically demonstrated that transverse diffusion stabilizes the LSs in the wide aperture bidirectional class-A laser model of [5]. The stabilization is twofold: Diffusion reduces the parameter region of pulsing LSs allowing for the existence of these structures even in the isotropic cavity; and converts the LSs in true CSs, as the LSs in an ensemble can be written/erased independently.

**Acknowledgments**

This work has been supported by the Spanish MEC, MICINN and the European Union FEDER through Projects FIS2005-07931-C03 and FIS2008-06024-C03. K.S. acknowledges financial support from Generalitat Valenciana through the program *Ajudes per a estades d'investigadors invitats en centres de la Comunitat Valenciana*.